\documentclass[aip,amsmath,amssymb,reprint]{revtex4-2}

\usepackage{graphicx}
\usepackage{dcolumn}
\usepackage{bm}
\usepackage{float}
\usepackage{color}
\usepackage{xcolor}
\usepackage{amsmath}
\usepackage{amssymb}
\usepackage{soul}
\usepackage{titlesec}
\usepackage[nottoc]{tocbibind}
\usepackage[normalem]{ulem}

\usepackage[utf8]{inputenc}
\usepackage[T1]{fontenc}
\usepackage{mathptmx}

\begin{document}


\title{Dynamic Triad Interactions and Evolving Turbulence - Part 1 Theory: 4D Modal Interactions} 



\author{Clara M. Velte}
\email[]{cmve@dtu.dk}
\homepage[Personal web page: ]{https://www.staff.dtu.dk/cmve}
\homepage[Projects web page: ]{https://www.trl.mek.dtu.dk/}
\affiliation{Department of Civil and Mechanical Engineering, Technical University of Denmark, Koppels All\'{e}, Building 403, 2800 Kongens Lyngby, Denmark}

\author{Preben Buchhave}
\email[]{buchhavepreben@gmail.com}
\affiliation{Intarsia Optics, S{\o}nderskovvej 3, 3460 Birker{\o}d, Denmark}

\date{\today}

\begin{abstract}
We investigate the effect of a four-dimensional Fourier transform on the formulation of the Navier-Stokes equation in Fourier space and the way the energy is transferred between Fourier components. Since time in a sampled high intensity turbulence must be considered a stochastic variable in the energy exchange between scales, we refer to these dynamic triad interactions as modal interactions, rather than the commonly referred to triad interactions in the classical 3-dimensional analysis. The inclusion of time as a parameter broadens the phase match condition from the classical one, $\Delta \bm{k} \cdot \bm{r} = \left [ \bm{k} - (\bm{k}_1 + \bm{k}_2 ) \right ] \cdot \bm{r}$, to the more general formulation that also includes temporal frequencies: $\Delta \bm{k} \cdot \bm{r} - \Delta \omega t = \left [ \bm{k} - (\bm{k}_1 + \bm{k}_2 ) \right ] \cdot \bm{r} - \left [ \omega - \left (\omega_1 + \omega_2 \right ) \right ] t$. This renders possible the occurrence of `delayed' and `advanced' interactions. The observation that mismatches in the wavevector triadic interactions may be compensated by a corresponding mismatch in the frequencies supports the empirically deduced delayed interactions reported in [Josserand \textit{et al.}, \textit{J. Stat. Phys.} (2017)]. These results explain the occurrence and inherent time development of the so-called Richardson cascade and also how finite temporal overlap of wave components can result in significant non-local interactions and consequently non-equilibrium turbulence, e.g., fractal grid generated turbulence. The consequences of including time as a parameter in practical experiments or simulations in terms of limited resolution, domain size etc. are treated in the companion paper (Part 2) of the present work. 
\end{abstract}

\maketitle

\section{\label{sec:1}Introduction}

In fluid mechanics and in particular in the theory of turbulence, the term triad interaction refers to the elementary momentum interchanges between Fourier components (or other basis functions) of the velocity field in wave vector space. As is generally accepted, the time evolution of all non-pathological flows is described by the Navier-Stokes equation, which is a second order partial differential equation with one nonlinear term, the so-called convective (or advective) term. The nonlinear term in the Navier-Stokes equation is of the $2^{nd}$ order and allows only two Fourier components to combine at a time to form a third one. The study of these triad interactions reveals the inner workings of the evolution of the turbulent velocity field and exposes the dynamics of the formation of the velocity structure and power spectra.

Most of the literature on the subject of triad interactions, see for example~\cite{Batchelorbook,4,5,6}, describes the ideal case of a homogeneous flow with infinite spatial ranges. Although time as such is not explicitly considered as a parameter in these classical descriptions, the spatial coordinates, $\bm{r} = (x,y,z)$ and the temporal coordinate, $t$, would then be coupled through the velocity at each point in the fluid, $d\bm{r} = \bm{u}(\bm{r},t)\, dt$, and, ideally, as the Navier-Stokes equation is a fully deterministic equation, there would be no need to describe the velocity as a stochastic variable. However, as is well known, it is beyond present-day computational capabilities to solve the Navier-Stokes equation for turbulence with a realistic span of spatial scales such that the space-time coupling at a point in $d\bm{r} = \bm{u}(\bm{r},t)\, dt$ is preserved, and hence we resort to a statistical description~\cite{Part2}. Furthermore, we want to study the turbulent mode interactions in realistic experimental situations or numerical simulations, which means that we will not be able to measure the velocity without a certain amount of uncertainty or noise. We must then consider a flow with a finite spatial range and a finite temporal range sampled with a finite sampling rate. Moreover, the finite spatial resolution of the measurement volume decouples the spatial and temporal fluctuations of the velocity and forces us to consider the velocity as a stochastic function of four independent parameters, the three spatial coordinates and time~\cite{iTi2021}. These aspects of practical experiments are the focus of the companion paper to the current work~\cite{Part2}, whereas the present work focuses on the consequences of including time as a parameter in the analysis of triad interactions. Temporal fluctuations of the velocity have not traditionally been considered in the study of triad interactions. The purpose of this work is, first and foremost, to try to understand the interactions that take place between velocity structures of different scales and to understand the underlying physical processes.

It is important in this work to distinguish between the physical velocity that in case of high intensity turbulence consists of velocity structures that fluctuate as a function of space and time, and on the other hand, the Fourier components that are mathematical plane waves extending over all space and time and whose size and number are defined by us mathematically. In physical space, a velocity wave structure cannot retain its shape for long -- the high-velocity parts will catch up with the low-velocity parts, the result being a sharpening of the wave front and a generation of higher and higher harmonic frequencies~\cite{2}. The tendency of this process to lead to a shock formation is prevented by the dissipation, which removes energy preferentially from the highest frequencies.

By including the Fourier transform of the fourth variable, namely time, in the Navier-Stokes equation, we have succeeded in deriving an expression for the efficiency of triad interactions, be it local or non-local, and in finding expressions for the time delays in the turbulent cascade process. The time evolution of the spectrum has previously been studied by Josserand \textit{et al.}~\cite{3}, who by time correlation methods found cascade delays depending both on the velocity magnitude and on the distance between Fourier components. Other attempts to derive the importance of local versus nonlocal triad interactions have considered only spatial triad interactions~\cite{4}. The Richardson cascade~\cite{5} and the model for the energy transfer in local equilibrium turbulence~\cite{Batchelorbook} put forth by Kolmogorov~\cite{6} presumed predominantly local interactions. The question if the energy cascade is primarily caused by local interactions was already discussed in Kraichnan~\cite{4} who pointed out the importance of vortex filaments and sheets in non-local interactions and the difficulty for the Kolmogorov theory to include the effect of the large structures. The issue was addressed by Zhou~\cite{9} and Grimaji and Zhou~\cite{10} and recently by analyzing direct numerical simulations (DNS)~\cite{11,12,13}.

Fourier decomposition is a familiar and fundamental tool for the description of spectral properties and has the advantages of descriptions using analytical functions with well-defined wavenumbers and frequencies. However, other bases may for some purposes be better suited to fluid turbulence such as interactions between helical modes or those that are optimal in the sense of expansion in, e.g., energy eigen-modes using proper orthogonal decomposition (POD)~\cite{14,15,16,17,17a,17b,SPODvsFT}. Although harmonic functions can only be energy-optimal in general on domains that are both periodic and homogeneous~\cite{SPODvsFT}, we use Fourier analysis according to common convention, since it provides distinct wavenumbers/frequencies and as a familiar conceptual tool only to illustrate the resulting processes without implications on the resulting analysis. We concentrate on the four-dimensional Fourier decomposition in a finite spatial and temporal range without loss of generality of the derived results.

That time should be included in the description of the turbulent field in order to capture the full four-dimensional fluctuations of the velocity was already pointed out by W. K. George~\cite{18}. In the present work, recognizing the fact that turbulence is a function of both three spatial coordinates and time, we find that the time fluctuations only have a direct effect on the triad interactions in the case of a finite measurement volume and high intensity turbulence. When Taylor's Hypothesis can be invoked, the physical velocity at a measurement point leads to a phase factor increasing linearly with time so the Fourier modes may be considered travelling waves. In this case, the time dependence falls out of the analysis, since no temporal development of the turbulent flow can take place in an effectively frozen field.

Although much literature deals with traveling waves such as oceanographic waves and coastal waves and connections between their spatial and temporal behavior (i.e., dispersion relations), not much literature can be found on turbulent structures decomposed as traveling waves and possible dispersion relations in turbulent velocity fields.

We begin by exploring the four-dimensional travelling wave Fourier transform. We explore the effect of the phase term to include both space and time, how this leads to a broadening of the phase match condition and how this opens up the possibility for the occurrence of `delayed' and `advanced' interactions. We then discuss the consequences in terms of being able to explain temporal development of turbulent flows and how finite temporal overlap can help in understanding how significant non-local interactions and thus non-equilibrium turbulence may arise.

\section{\label{sec:2}The Navier-Stokes equation in 4-dimensional Fourier space}

It is generally accepted that most constant density and constant viscosity fluid flows are fully described by the Navier-Stokes equation. The Navier-Stokes equation is a momentum conservation equation valid for a continuous, isotropic fluid wherein the Newtonian deformation law reduces to the hydrostatic pressure for zero strain rates, e.g. when the fluid is at rest. The fact that we focus on the Navier-Stokes equation does not make our results and conclusions less general, since the focus is on the nonlinear mechanisms, but it simplifies the description of the coupling between shear stresses and strain rates in turbulence modeling.

The Navier-Stokes equation describes the fluid momentum in an infinitesimal control volume in physical space and time. In contrast, interactions between different scales are most often described in three-dimensional spatial Fourier space~\cite{19,20} using plane waves that extend throughout space. The nonlinear development of velocity fluctuations is thus often better described by considering the Navier-Stokes equation in the Fourier domain. Here, we generalize by including time as an independent variable in the Fourier transform and by limiting the analyzed flows in space and time. With this analysis, it becomes clear which frequency components interact with each other, what the efficiencies are of these interactions and how they influence the time development of the statistical properties of the turbulent flow, for example the time development of the spatial and temporal velocity power spectra.

The Navier-Stokes equation in the physical space coordinate system, normalized by setting the density equal to unity, is given by 
\begin{equation}
\frac{\partial \bm{u}}{\partial t} + \left ( \bm{u} \cdot \nabla \right ) \bm{u} = - \nabla p + \nu \nabla^2 \bm{u}
\end{equation}
which expresses the momentum change of a fluid  element with velocity $\bm{u}(\bm{r},t) = (u_1(\bm{r},t), u_2(\bm{r},t), u_3(\bm{r},t))$ where the position vector is given by $\bm{r} = (r_1, r_2, r_3)$ and time is denoted by $t$. The second term on the left-hand side of the normalized equation represents convective acceleration. The first term on the right-hand side represents the fluctuating force due to pressure and the second term represents the viscous action due to momentum diffusion. We consider constant density flows, which means that the divergence of the velocity field equals zero:
\begin{equation}
\nabla \cdot \bm{u} = 0
\end{equation}

We can re-write the Navier-Stokes equation in physical space, emphasizing the nonlinear term $\bm{N}(\bm{r},t)$:
\begin{equation}
\frac{\partial \bm{u}}{\partial t}\left ( \bm{r},t \right ) + \bm{N}\left ( \bm{r},t \right ) = - \nabla p\left ( \bm{r},t \right ) + \nu \nabla^2 \bm{u}\left ( \bm{r},t \right ).
\end{equation}
In Fourier space, expanding the classical purely spatial Fourier transform into covering not only all three spatial dimensions (as done e.g. by Kraichnan\cite{BobKraichnan} and Lee\cite{Lee1977}), but \textit{also} time, the equation becomes:
\begin{equation}\label{eq:NSF}
-i\omega \bm{\hat{u}}\left ( \bm{k},\omega \right ) + \bm{\hat{N}}\left ( \bm{k},\omega \right ) = -i\bm{k} \hat{p}\left ( \bm{k},\omega \right ) - \nu k^2 \bm{\hat{u}}\left ( \bm{k},\omega \right )
\end{equation}
where $k = |\bm{k}|$ for the wave vector $\bm{k} = (k_1, k_2, k_3)$, $\omega$ denotes temporal frequency and $\bm{\hat{N}}\left ( \bm{k},\omega \right )$ is the Fourier transformed nonlinear term and $\hat{p}$ the Fourier transform of the pressure.

Pressure can be eliminated by using the incompressibility condition in Fourier space
\begin{equation}
\bm{k}\cdot \bm{\hat{u}} \left ( \bm{k},\omega \right )= 0
\end{equation}
to find pressure by solving a Poisson equation over all space. When doing that, we see that pressure enters the Navier-Stokes equation in a complicated way, but that the mode interactions are still governed by the same nonlinear term. Taking the divergence of the Navier-Stokes equation~(\ref{eq:NSF}), we obtain
\begin{equation}
-i\omega \bm{k}\cdot \bm{\hat{u}}\left ( \bm{k},\omega \right ) + \bm{k} \cdot \bm{\hat{N}}\left ( \bm{k},\omega \right ) = -i k^2 \hat{p }\left ( \bm{k},\omega \right ) - \nu k^2 \bm{k}\cdot \bm{\hat{u}}\left ( \bm{k},\omega \right )
\end{equation}
where the first and last terms are zero due to continuity. Then
\begin{equation}
\hat{p}\left ( \bm{k},\omega \right ) = \frac{i}{k^2} \, \bm{k}\cdot \bm{\hat{N}}\left ( \bm{k},\omega \right )
\end{equation}
Inserting the above into~(\ref{eq:NSF}), we obtain
\begin{equation}
-i\omega \bm{\hat{u}}\left ( \bm{k},\omega \right ) + \left [ 1 - \frac{\bm{k}}{k^2} \, \left ( \bm{k}\cdot \right )\right ] \bm{\hat{N}}\left ( \bm{k},\omega \right )  =  - \nu k^2 \bm{\hat{u}}\left ( \bm{k},\omega \right )
\end{equation}
where $\frac{\bm{k}}{k^2} \, \left ( \bm{k}\cdot \right )$ is an operator applied onto $\bm{\hat{N}}\left ( \bm{k},\omega \right )$. The first term in the square bracket on the left side is due to convective acceleration and the second one is due to the pressure gradient term, $\nabla p$. It is important to note here, that the effect on both convection and the pressure gradient is governed by the same nonlinear term, $\bm{\hat{N}}(\bm{k},\omega)$.

Thus, assuming the velocity vector is known, either through measurement, modeling or solution of Poisson's equation with the given boundary conditions, the modal interactions between Fourier components are determined by the nonlinear term $\bm{\hat{N}}(\bm{k},\omega)$. In the following, we therefore proceed by analyzing the effect of this term, assuming that the velocity is a stochastic function of four parameters, namely 3-dimensional space and time.\\

\section{\label{subsec:2.1}4-dimensional Fourier decomposition with travelling waves}

As we want to study the nonlinear interactions of the velocity structures, we limit our analysis to the nonlinear term in the Navier-Stokes equation. We will herein reserve the term `triad interactions' to the interactions between spatial Fourier components, while the term `modal interactions' will be used for the more general interactions across wavenumber and frequency.

The 4-dimensional Fourier transform over all space and time is given by
\begin{eqnarray}
\bm{\hat{u}}(\bm{k},\omega) = \iiiint\limits_{-\infty}^{\infty} \bm{W}(\bm{r}) W(t)\bm{u}(\bm{r},t) e^{-i(\bm{k}\cdot \bm{r}-\omega t)}\, d\bm{r} \, dt
\end{eqnarray}
where $\bm{W}(\bm{r})$ and $W(t)$ are window functions (e.g. rectangular) delimiting a finite region of an otherwise homogeneous (infinite) velocity field and time record. The presence of the window functions reflects the physical reality of a finite region of the velocity field and eliminates the problem of an infinite integral and infinite energy. As is well known, a rectangular window will result in a sinc-squared window in the spectral domain. These matters will be further discussed in the companion paper~\cite{Part2}.

Expanding the velocity field in 4-dimensional Fourier components amounts to an expansion in travelling plane waves (which includes the conventional fixed spatial wave field for $t = \mathrm{constant}$):
\begin{eqnarray}\label{eq:3}
\bm{u}(\bm{r},t) = \frac{1}{(2\pi)^4}\iiiint\limits_{-\infty}^{\infty} e^{i(\bm{k}\cdot \bm{r}-\omega t)}\bm{\hat{u}}(\bm{k},\omega)\, d\bm{k} \, d\omega
\end{eqnarray}
This equation represents the spectral decomposition of the velocity field with $\bm{\hat{u}}(\bm{k},\omega)$ being the velocity Fourier components of the spatially and temporally limited field.

Fourier waves are abstract mathematical concepts, and physical waves are built up by superposition knowing the values of the Fourier coefficients. Thus, also the triad interactions are mathematical concepts. Even so, the strengths and phase relations between the Fourier coefficients $\bm{\hat{u}}(\bm{k},\omega)$ do provide invaluable information about the development of the real physical velocity field, $\bm{u}(\bm{r},t)$.

Now, consider the spectral representation of the nonlinear convection term, $\bm{\hat{N}}(\bm{k},\omega) = FT \{ \left ( \bm{u}(\bm{r},t) \cdot \nabla \right )\bm{u} (\bm{r},t)\}$:
\begin{widetext}
\begin{eqnarray}\label{eq:4}
  \bm{\hat{N}}(\bm{k},\omega) &=& \iiiint\limits_{-\infty}^{\infty} \bm{W}(\bm{r}) W(t) e^{-i(\bm{k}\cdot \bm{r}-\omega t)}\, d\bm{r} \, dt
\times \nonumber\\
&&\left [ \left ( \frac{1}{(2\pi)^4} \iiiint\limits_{-\infty}^{\infty} e^{i(\bm{k}_1\cdot \bm{r}-\omega_1 t)}\bm{\hat{u}}(\bm{k}_1,\omega_1) \, d\bm{k}_1 \, d\omega_1 \cdot \nabla \right )
\frac{1}{(2\pi)^4} \iiiint\limits_{-\infty}^{\infty} e^{i(\bm{k}_2\cdot \bm{r}-\omega_2 t)}\bm{\hat{u}}(\bm{k}_2,\omega_2) \, d\bm{k}_2 \, d\omega_2 \right ]
\end{eqnarray}
\end{widetext}
Evaluating the spatial gradients, we obtain:
\begin{widetext}
\begin{eqnarray}\label{eq:5}
  \bm{\hat{N}}(\bm{k},\omega) &=& \iiiint\limits_{-\infty}^{\infty} \bm{W}(\bm{r}) W(t) e^{-i(\bm{k}\cdot \bm{r}-\omega t)}\, d\bm{r} \, dt
\times \nonumber\\
&&\left [ \left ( \frac{1}{(2\pi)^4} \iiiint\limits_{-\infty}^{\infty} e^{i(\bm{k}_1\cdot \bm{r}-\omega_1 t)}\bm{\hat{u}}(\bm{k}_1,\omega_1) \, d\bm{k}_1 \, d\omega_1 \right )
\frac{i}{(2\pi)^4} \iiiint\limits_{-\infty}^{\infty} e^{i(\bm{k}_2\cdot \bm{r}-\omega_2 t)} \, \bm{k}_2 \cdot \bm{\hat{u}}(\bm{k}_2,\omega_2) \, d\bm{k}_2 \, d\omega_2 \right ]
\end{eqnarray}
\end{widetext}
Changing the order of integration and collecting the exponentials, we find:
\begin{widetext}
\begin{eqnarray}\label{eq:6}
  \bm{\hat{N}}(\bm{k},\omega) &=& \frac{1}{(2\pi)^4} \iiiint\limits_{-\infty}^{\infty} \, d\bm{k}_1 \, d\omega_1 \frac{1}{(2\pi)^4} \iiiint\limits_{-\infty}^{\infty} \, d\bm{k}_2 \, d\omega_2 \times \nonumber\\
  && \iiiint\limits_{-\infty}^{\infty} \bm{W}(\bm{r}) W(t) e^{-i[(\bm{k}-\bm{k}_1-\bm{k}_2)\cdot \bm{r}-(\omega -\omega_1 - \omega_2) t]}\, d\bm{r} \, dt \left [ \left ( i\bm{k}_2 \cdot \bm{\hat{u}}(\bm{k}_1,\omega_1) \right ) \bm{\hat{u}}(\bm{k}_2,\omega_2) \right ]
\end{eqnarray}
\end{widetext}
This general phase match condition means that phase match can be achieved with data recorded at different times in combination with a certain spatial position, hence there may exist `delayed interactions' and `advanced interactions'. However, it must be understood that due to the fundamental second order non-linear term in the Navier-Stokes equation, the interaction is still between three entities, i.e. a triadic interaction.

Let us analyze the phase of the exponential in the $(\bm{r},t)$-integral in equation~(\ref{eq:6}) in further detail.

\section{\label{subsec:2.2}The classical phase match and Taylor's hypothesis}
As in the special case of the classical approach, the nonlinear term in equation~(\ref{eq:6}) could be phase matched if both $\bm{k}=\bm{k}_1 + \bm{k}_2$ and $\omega = \omega_1 + \omega_2$. If the record tends to infinity, the interactions are limited to the product of two delta functions:
\begin{equation}\label{eq:7}
\delta(\bm{k},\omega) = \delta (\bm{k} - \bm{k}_1 - \bm{k}_2) \delta (\omega - \omega_1 - \omega_2)
\end{equation}
In this special case, both the wave vectors and the temporal frequencies must be phase matched independently for efficient interactions to occur.

If we can invoke Taylor's Hypothesis during some finite extent in time and space, e.g. when the turbulence intensity is less than approximately 20\%, the time fluctuations and spatial fluctuations are coupled as $\bm{u}_0 = \bm{u}(\bm{r}_0,t_0)$, where $\bm{u}_0$ is the convection velocity of the frozen turbulence through the control volume. The spatial pattern now moves with a constant velocity $\bm{u}_0$ through the control volume, and the time integral leads to the same phase match condition as the spatial one: $\left [ \omega - (\omega_1 + \omega_2)\right ] t \Rightarrow \left [ \bm{k} - (\bm{k}_1 + \bm{k}_2) \right ]\cdot \bm{u}_0$. Consequently, time does not play an independent role in this case.

In the case of infinite records, the phase match conditions for the spatial and temporal frequencies are:
\begin{widetext}
\begin{eqnarray}\label{eq:8}
\bm{\hat{N}}(\bm{k},\omega) &=& \frac{1}{(2\pi)^4}\int_{\omega_2} \iiint_{\bm{k}_2} \, d\bm{k}_2 \, d\omega_2 \left [ \left ( i\bm{k}_2 \cdot \bm{\hat{u}}(\bm{k} - \bm{k}_2,\omega - \omega_2) \right ) \bm{\hat{u}}(\bm{k}_2,\omega_2) \right ]
\end{eqnarray}
\end{widetext}
Thus, when we can invoke Taylor's Hypothesis, we can argue that at a time later, $t_0+t$, the spatial Fourier transform is unchanged except for a phase factor:
\begin{eqnarray}
\bm{\hat{u}}(\bm{k},t) = e^{i  \bm{k}\cdot \bm{u}_0 t}\bm{\hat{u}}(\bm{k},t_0), \nonumber
\end{eqnarray}
assuming $\bm{u}_0 = \bm{u}(\bm{u}_0,t_0)$ nearly constant.

Thus, the Fourier components form travelling waves:
\begin{eqnarray}
\bm{\hat{u}}(\bm{k},t) = \iiint \bm{W}(\bm{r})e^{-i \bm{k}\cdot (\bm{r}-\bm{u}_0 t)} \bm{u}(\bm{r},t_0) \, d\bm{r} \nonumber
\end{eqnarray}
and the nonlinear term becomes
\begin{equation}\label{eq:9}
\bm{\hat{N}}(\bm{k}) = \iiint_{\bm{k}_2} \left [ i \bm{k}_2\cdot \bm{\hat{u}}(\bm{k}-\bm{k}_2, (\bm{k}-\bm{k}_2)\cdot \bm{u}_0)\right ] \bm{\hat{u}}(\bm{k}_2,\bm{k}_2\cdot \bm{u}_0) \, d\bm{k}_2
\end{equation}
This is the conventional expression for spatial delta-function triad interactions assuming an infinite spatial domain.

\section{\label{subsec:2.3}Instant vs delayed interactions}
If we instead consider the total exponent in equation~(\ref{eq:6}), 
$$\Delta \bm{k} \cdot \bm{r} - \Delta \omega t = \left [ \bm{k} - (\bm{k}_1 + \bm{k}_2 ) \right ] \cdot \bm{r} - \left [ \omega - \left (\omega_1 + \omega_2 \right ) \right ] t$$
we see that a mismatch in the spatial phase, $\Delta \bm{ k}\cdot \bm{r} = \left [ \bm{k} - \left ( \bm{k}_1 + \bm{k}_2 \right ) \right ] \cdot \bm{r}$, may be compensated by a temporal phase mismatch, $\Delta \omega t = \left [ \omega - \left ( \omega_1 + \omega_2 \right ) \right ]t$. In order to understand what this implies, we note that the nonlinear term, $\bm{N}(\bm{r},t)$, is evaluated by the Navier-Stokes equation at a particular location and at a particular time. The variables $\bm{r}$ and $t$ in equation~(\ref{eq:6}) refer to the complete velocity record in all space and time, which must be known in order to evaluate the Fourier transform. Thus the triade contributions related to the complete exponent in equation~(\ref{eq:6}) may be generated from velocities at times prior to or after the current time, where the Navier-Stokes equation operates. These dynamic triad interaction contributions we may denote ``advanced'' and ``delayed'' modal interactions.

However, as $\bm{r}$ and $t$ in the Fourier transform exponent in equation~(\ref{eq:6}) vary independently, there will not be a coherent phase match for the whole integral. Instead we will see spurious positive and negative contributions, which will appear as a broad band turbulence pattern moving across space as travelling waves as $t$ in equation~(\ref{eq:6}) increases.

We can conclude that when the turbulence intensity is so high that we cannot speak of a fixed pattern of velocity convecting past a point in space, the ``triad interactions'' are in reality four-dimensional modal interactions. This implies a broadening of the possible interactions and the possibility of delayed interactions between a range of different spatial and temporal Fourier components, since the temporal phase shift is time dependent.

\section{Discussion}

All this has interesting implications for high intensity turbulent flows where Taylor's Hypothesis cannot be invoked. First, it explains the inherent delay in development of turbulence and energy exchange between scales (e.g. the Richardson cascade). An example of a (slowly) developing turbulent flow is the classical decaying grid generated turbulence. Second, as a consequence of introducing time as an independent variable in the governing equations, the energy exchange of travelling waves will not only depend on the spatial overlap of the two waves. The finite temporal overlap can also have a significant impact on the resulting energy exchange between the two waves. Consequently, waves with very different wavelengths travelling at the same velocity (speed and direction) can have significant tempo-spatial overlap, hence giving rise to significant non-local interactions. Prominent examples of such non-equilibrium turbulent flows are homogeneous shear flow turbulence, decaying turbulence and fractal grid generated turbulence, c.f. George~\cite{ReconsideringK41}. The same kinds of effects may even occur in what is commonly considered equilibrium flows, e.g. thin shear flows (boundary layers, jets, wakes, mixing layers etc.), or in the vicinity of large vortices. These aspects are further discussed in the companion paper to the current work~\cite{Part2}. In this companion paper\cite{Part2}, we present examples of such modal interactions in real flows such as flows with a velocity dominated by a few, strong oscillatory modes and also an example of a fully turbulent, high intensity developing flow.

\section{\label{sec:6}Conclusions}

We studied the modal interactions between the Fourier components of a full four-dimensional Fourier transform of a turbulent velocity field including both the three spatial coordinates and time. We have analyzed various forms of the resulting phase factor in the 4-dimensional Fourier transform and the effect on the modal interactions. In this process, we found that time must be considered an independent variable when Taylor's frozen field hypothesis does not apply or when the finite size probe volume decouples time and space in the otherwise valid definition $\bm{u}(\bm{r},t) = d\bm{r} / dt$.

In cases where Taylor's hypothesis is not applicable, the fluctuating temporal phase can compensate for a spatial phase mismatch and through that effect broaden the phase match condition. This can result in `delayed' and `advanced' interactions and it also explains the inherent time development in developing turbulent flows, including the time required for the cascade development in the Richardson cascade. The results can also explain how the finite temporal overlap of wave components can yield significant non-local interactions, which can be used to explain the varying development times for e.g. classical decaying grid turbulence and the highly non-equilibrium flow produced by fractal grid generated turbulence (which development length is typically an order or two of magnitude shorter than for classical grid generated turbulence).  

These investigations have resulted in a better understanding of modal interactions and illustrated the influence of stochastic time fluctuations.

\begin{acknowledgments}
The authors wish to thank Professor Emeritus Poul Scheel Larsen for many helpful discussions. 
The final stages of the project would not have been possible without the support of CV by the European Research Council and of CV and PB by the Poul Due Jensen Foundation: 
This project has received funding from the European Research Council (ERC) under the European Union's Horizon 2020 research and innovation programme (grant agreement No 803419). 
Financial support from the Poul Due Jensen Foundation (Grundfos Foundation) for this research is gratefully acknowledged.
\end{acknowledgments}

\bibliography{aiptemplate_NStheory}

\end{document}